\documentclass[aps,twocolumn,superscriptaddress,prb,amsmath,amssymb,showpacs,lengthcheck,reprint,floatfix]{revtex4-1} %revtex4-1
\usepackage[T1]{fontenc}
\usepackage[latin9]{inputenc}
\usepackage{mathrsfs}
\usepackage{graphicx}
\usepackage[usenames,dvipsnames]{color}

			    \definecolor{mygray}{gray}{0.88}
\usepackage{braket}
\usepackage{placeins}
\newcommand\ret{r}

\usepackage{blindtext}

\begin{document}

\title{Carrier dynamics in quantum-dot tunnel-injection structures: microscopic theory and experiment}  
\author{Michael Lorke}
\affiliation{Institute for Theoretical Physics, University of Bremen, Otto-Hahn-Allee 1,  Bremen, 28359,  Germany}
\author{Igor Khanonkin}
\affiliation{Electrical Engineering Department and Russel Berrie Nanotechnology Institute, Technion, Haifa, 32000, Israel}
\author{Stephan Michael}
\affiliation{Institute for Theoretical Physics, University of Bremen, Otto-Hahn-Allee 1, Bremen, 28359, Germany}
\author{Johann Peter Reithmaier}
\affiliation{Technische Physik, Institute of Nanostructure Technologies and Analytics, Center of Interdisciplinary Nanostructure Science and Technology (CINSaT), University of Kassel, Kassel, 34132, Germany}
\author{Gadi Eisenstein}
\affiliation{Electrical Engineering Department and Russel Berrie Nanotechnology Institute, Technion, Haifa, 32000, Israel}
\author{Frank Jahnke}
\affiliation{Institute for Theoretical Physics, University of Bremen, Otto-Hahn-Allee 1, Bremen, 28359, Germany}
\begin{abstract}

Tunneling-injection structures are incorporated in semiconductor lasers in order to overcome the fundamental dynamical 
limitation due to hot carrier injection by providing a carrier transport path from a cold carrier reservoir. 
The tunneling process itself depends on  band alignment between quantum-dot levels and the injector quantum well,
especially as in these devices LO-phonon scattering is dominant.
% As inhomogeneous broadening is omnipresent in QD structures, individual members of the ensemble couple differently to the injector well.
% 
Quantum dots with their first excited state near the quantum well bottom profit most from  tunnel coupling.
As inhomogeneous broadening is omnipresent in quantum dot structures, this implies that individual members of the ensemble couple differently to the injector quantum well.
Quantum dots with higher energy profit less, as the phonon couples to higher, less occupied states.
Likewise, if the energy difference between ground state and quantum well exceeds the LO phonon energy, 
scattering becomes increasingly inefficient.
Therefore, within 20-30meV we find Quantum Dots that benefit substantially different from the tunnel coupling.
Furthermore, in quantum dots with increasing confinement depth, excited states become sucessively confined.
Here, scattering gets more efficient again, as subsequent excited states  
reach the phonon resonance with the quantum well bottom.
Our results provide guidelines for the optimization of tunnel-injection lasers.
% Hence, we observe an alternating behavior of efficient and inefficient scattering over the inhomogeneous ensemble.
%
Theoretical results for electronic state caluluations in connection with carrier-phonon and carrier-carrier scattering are compared to experimental results of the temporal gain recovery after a short pulse perturbation.

% In quantum dots with increasing confinement depth,  excited states become sucessively confined.
% 
% For those members with low ground state energies, that get out of resonance with the injector quantum well, scattering becomes increasingly inefficient.
% 
% For even lower ground state energies, scattering gets more efficient again, as the next excited state becomes 
% confined and comes \emph{into} resonance.
% 
% Hence, we observe an alternating behavior of efficient and inefficient scattering over the inhomogeneous ensemble.
% 
% This is analyzed in conjunction with new experiment via the temporal gain response to a short pulse perturbation and demonstrated in its effect on the gain recovery.

% Their design depends on careful alignment of quantum-dot levels with the injector quantum well,
% especially as in these devices LO-phonon scattering is dominant, which is more alignment dependent than Coulomb scattering. 
% 
% 
% As inhomogeneous broadening is omnipresent in QD structures, different members of the inhomogeneous ensemble couple differently to the injector well
% 
% 
% While for shallow dots the 
% 
% At a certain point a higher excited states becomnes localized and sc
% 
% 
% 
% and
% we identify alternating regimes of scattering efficiency. 
% The tunneling process itself depends on energy band alignment and various operating conditions. 
% A microscopic theoretical analysis of gain dynamics in tunneling-injection 
% devices is used to studz the efficiency of tunnel injection processes in conjunction with new experiments. 

\end{abstract}
\maketitle

% \begin{figure}[h!]
%   \begin{center}
% %   \hspace*{-2cm}%
%     \includegraphics[trim=2.5cm 15.5cm 2.8cm 3.3cm,clip,width=0.4\textwidth,angle=0]{./TI_laser_theory.pdf}
%     \caption{Building blocks of the semiconductor laser theory.
%     \label{schematic1}}
%   \end{center}
% \end{figure}

\section{Introduction}

One of the basic physical mechanisms limiting the modulation capabilities of semiconductor lasers is the gain 
nonlinearity which is often driven by hot carrier injection \cite{Coldren:95}. An elegant scheme to 
tackle this hurdle is to employ tunneling injection (TI) where cold (low-energetic) carriers from an injector quantum well (iQW) 
reservoir are fed directly to the lasing state \cite{Khanonkin2022}.
The successful use of TI was demonstrated for quantum well lasers more than twenty years ago   \cite{Sun:93} and later for
quantum dot (QD) lasers at short wavelengths \cite{Bhattacharya:92} as well as at 1550 nm \cite{Bhowmick:14}. For the QD
lasers, TI was also found to improve the temperature stability \cite{Han:08}.

The tunnel barrier is used  for controlling the coupling of QDs to an 
iQW thereby tailoring the carrier transport into the laser levels by means of carrier scattering processes. For the 
description of carrier scattering through the nano-scale structure we use a 
quantum mechanical framework of the electronic states and interaction processes 
in a microscopic model \cite{Michael:18,lorke2018performance,Khanonkin:20}. 
The device optoelectronic properties are directly linked to the carrier dynamics through 
the optical gain and gain dynamics of the active material, which dictate, in turn, the laser modulation capabilities.

In this Letter we provide a combined experimental and theoretical study of the gain dynamics in TI-QD systems. In particular, 
we present a time domain description of the ground state population response to a short 
pulse perturbation. This response is directly related to the modulated capabilities of the TI-QD laser under 
strong stimulated emission conditions when the cold carrier injection dominates.

\section{Theory}
Our starting point is the electronic structure of the TI-QD system, which is designed for operating at an emission wavelength of about 1550nm \cite{Khanonkin2022}. 
Three-dimensional wave-functions are obtained using the discretized {\bf k $\cdot$ p} Hamiltonian as implemented in the Nextnano3 package \cite{nextnano3}.
These single-particle calculations are used to determine energy level structures and interaction matrix elements for the combined iQW-QD system.
In a second step, time domain calculations are performed for a system with stationary carrier injection, 
in combination with pump-probe excitation.
To simulate the gain dynamics, we utilize the 
semiconductor Bloch equations,
\begin{equation}
\label{eq:sbeEQ}
\begin{split}
\left [ i\hbar  \frac{d}{d t}  -  \varepsilon^{c}_{\alpha}(t)+ \varepsilon^{v}_{\alpha}(t)\right ]&\psi^{cv}_{\alpha}(t)\\
+\left[  f^v_{\alpha}(t)- f^c_{\alpha}(t) \right ]& \Omega^{cv}_{\alpha}(t) =-i\hbar S^{cv}_{\alpha}(t)~, \\[0.3cm]
i\hbar  \frac{d}{d t}f^c_{\alpha}(t)+\Omega^{cv}_{\alpha}(t)&\left[\psi^{cv}_{\alpha}(t)\right ]^\ast\\
- \psi^{cv}_{\alpha}(t)&\left[ \Omega^{cv}_{\alpha}(t)\right ]^\ast =-i\hbar S^{cc}_{\alpha}(t)~,
\end{split} 
\end{equation}
where $\varepsilon^{(c,v)}_{\alpha}$ are the Hartree-Fock (HF) renormalized energies of state $\alpha$ in the conduction (valence) band,
$f^{(c,v)}_{\alpha}$ are population functions, $\psi^{cv}_{\alpha}$ is the interband transition amplitude and $\Omega^{cv}_{\alpha}$ the (renormalized) Rabi energy.
These include the HF renormalizations of the single-particle energies and of the Rabi energy

\begin{align}\label{eq:hfren}
\nonumber \varepsilon^{c}_{\alpha}(t)&= e^{c}_{\alpha}(t) - \sum\limits_\beta 
V^\text{cccc}_{\alpha\beta\alpha\beta}f^{c}_\beta(t)\\
\nonumber \Omega^{cv}_{\alpha}(t)&= d^{cv}_{\alpha}E(t)(t) + \sum\limits_\beta 
V^\text{cvvc}_{\alpha\beta\alpha\beta}\psi^{cv}_\beta(t)~.
\end{align}

The scattering contributions $S_{\alpha}$ on the right-hand side of Eq.~\eqref{eq:sbeEQ} account for both carrier-phonon and carrier-carrier scattering.
We have shown in Ref.~\cite{Michael:18} that carrier-phonon scattering is the dominant interaction process in TI structures.
The corresponding scattering rates are described via generalized Boltzmann scattering rates including quasi-particle effects, 
taken from a microscopic polaron theory for TI and dot in a well (DWELL) structures \cite{Seebeck:05,Lorke:06},
\begin{equation}\label{eq:qk-scat}
\begin{split}
S^{cc}_\alpha = 2\text{ Re}\sum_{\beta}\int\limits^t_{-\infty}dt'~|M_{\alpha\beta}|^2
 &G^{\ret}_{\beta}(t-t')\left[G_{\alpha}^{\ret}(t-t')\right]^* \\
 \ast\, \Big\{\left[f_{\beta}(t')(1-f_{\alpha}(t'))\right] i\hbar \Big[&(1+n_\text{LO})e^{- i\omega_\text{LO}(t-t')}\\
& +n_\text{LO}e^{+ i\omega_\text{LO}(t-t')}\Big] \\
 -\left[f_{\alpha}(t')(1-f_{\beta}(t'))\right]  i\hbar \Big[&(1+n_\text{LO})e^{+ i\omega_\text{LO}(t-t')}\\
& +n_\text{LO}e^{- i\omega_\text{LO}(t-t')}\Big]\Big\}~.
\end{split}
\end{equation}
Here, $G^{\ret}_{\alpha}$ is the retarded polaronic Green's function for the state $\alpha$, $M_{\alpha\beta}$ are the interaction matrix elements as determined
from the {\bf k $\cdot$ p} wave functions and $n_\text{LO}$ is the phonon population. Details of the theoretical description are given in Ref.~\cite{Michael:18}.
Carrier-carrier scattering is incorporated within a relaxation time approximation \cite{Michael:18}.

\begin{figure}[h!]
  \begin{center}
%    \hspace*{2cm}%
    \includegraphics[width=0.6\textwidth,angle=0]{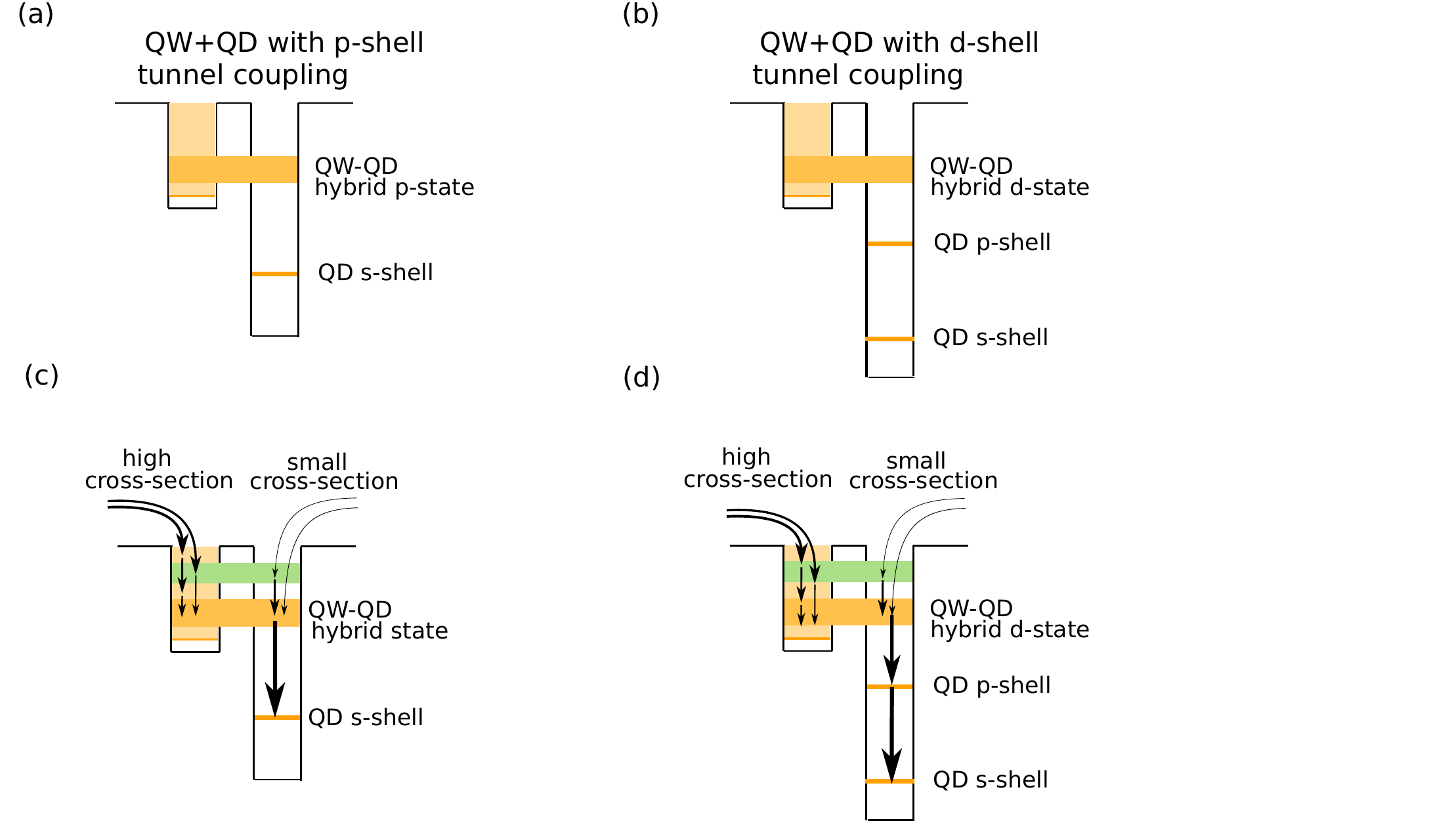}
    \caption{
(a) QD-QW system with hybridized p-shell. \newline
\,(b) QD-QW system with localized p-shell and hybridized d-shell. (c) and (d) Scattering processes corresponding to (a) and (b).
    \label{schematic2}}
  \end{center}
\end{figure}

We consider an inhomogeneous QD ensemble with size and composition fluctuations, leading to a Gaussian distribution of ground 
state (s-shell) energies.
Depending on the size and material composition of the QDs, we encounter two scenarios for carrier relaxation into the ground state.
\begin{enumerate}
 \item[a)] For QDs with a rather shallow confinement, the p-shell hybridizes with the injector QW. This corresponds to the 
 situation considered in our previous work\cite{Michael:18,lorke2018performance} and is shown in Fig. \ref{schematic2}a.
 \item[b)] For QDs with a deeper confinement, the p-shell becomes localized, as it does not overlap energetically with the QW. On the other hand, the d-shell 
 hybridizes efficiently with the iQW. This is shown in Fig. \ref{schematic2}b.
\end{enumerate}

Capturing of carriers from the barrier into the TI-QD system is dominated by the scattering into iQW states, due to its larger spatial cross-section compared to the QDs.
At the same time one can not distinguish  carrier capture via the iQW states from direct capture into higher QD states, since
these are joined states that are connected by the tunnel coupling, indicated by the orange and green shaded areas in Fig.~\ref{schematic2}c and d. 
Our approach takes care of this, as it considers scattering in the joined eigenstates of the iQW-QD system.

The scattering into the s-shell takes place via two different routes.
For the situation with a hybridized p-shell, Fig.~\ref{schematic2}a and c, the scattering is dominated by LO-phonon 
interaction, provided that energy seperation does not exceed the LO-phonon energy by more than 20\%. Otherwise, Coulomb scattering takes over, however, at a much weaker transition rate into 
the s-shell.
For the situation with a hybridized d-shell, LO-phonon scattering provides the dominant mechanism for scattering between the hybridized QW states and the localized QD p-shell.
Once carriers occupy any confined QD state, Coulomb interaction provides efficient intra-QD relaxation into the s-shell.

\section{Experiment}
\begin{figure}[h!]le
	\begin{center}
		\includegraphics[width=0.5\textwidth,angle=0]{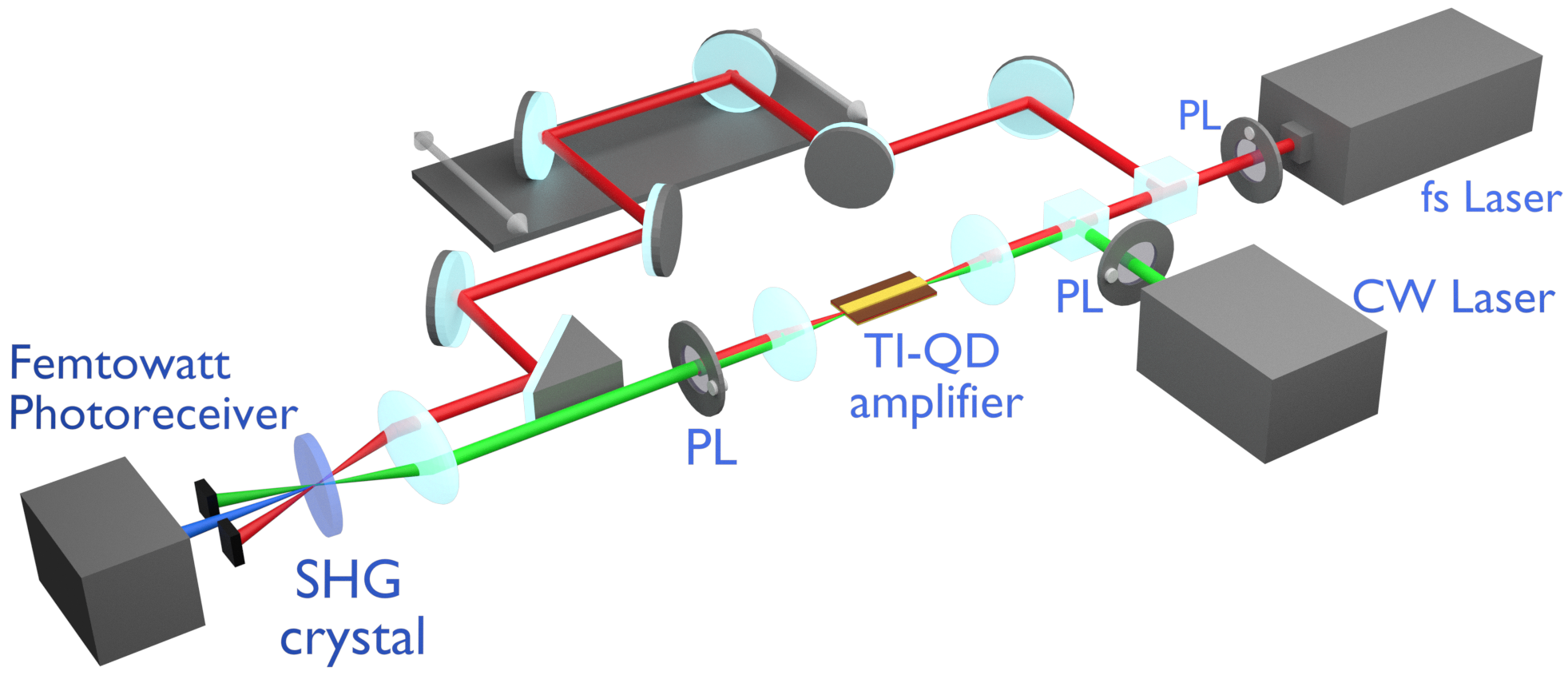}
		\caption{
			\label{setup} Multi-wavelength pump-probe system. PL stands for polarizer.}
	\end{center}
\end{figure}

Time-resolved gain recovery across the inhomogeneously broadened QD ensemble was measured using multi-wavelength 
pump-probe \cite{Khanonkin:20}, depicted in Fig. \ref{setup}. Near transform-limited 100 
fs laser pulses are split to form a pump and a reference, while a continuous wave signal generated by a tunable laser serves as the probe. 
The TM polarized pump is combined with the TE polarized probe and coupled to a TI-QD amplifier. Following the 
femto-second perturbation, the pump pulse is removed at the amplifier output by a polarizer. 
The output probe is sampled by the reference pulse in a non-linear crystal and its second harmonic is detected using a femtowatt photoreceiver. 
The optical delay between the pump and the reference pulses is scanned with femto-second resolution allowing to map out the 
gain dynamics with high temporal resolution and at any wavelength across the inhomogeneous QD gain spectrum.

The TI-QD sample comprised a 1 mm long ridge waveguide laser, whose facets were anti-reflection coated. The intrinsic layer 
of the TI-QD gain medium includes six QD-iQW pairs, where each pair is separated by  a 2 nm wide tunneling barrier layer. 
The QDs density was $3 \cdot 10^{10} cm^{-2}$. The ensemble exhibits high homogeneity with a  photo-luminescence line 
width of 22 meV at 10 K \cite{Bauer:21}.

\section{Results}
To investigate characteristic properties of the carrier relaxation from the injector QW states into the inhomogeneous ensemble 
of QD ground states, we examine the recovery of the differential pump-probe signal following a short pulse perturbation.
Our starting point of the calculations is a quasi-equilibrium population of the electrons in the combined QD and injector-QW system 
with an excited carrier density of  $10^{12}$ cm$^{-2}$.
A 100fs optical pulse, tuned to the center of the inhomogeneous QD distribution, is applied to partially deplete the QD ground states. 
The pulse fluence is chosen to ensure that the refilling of the QD states via the scattering 
is connected with only a very small drop in the excited QW carrier density. In other words, the refilling of the QD states is not 
limited by the available carriers in the injector QW. 

\begin{figure}[h!]
  \begin{center}
%   \hspace*{-2cm}%
    \includegraphics[width=0.5\textwidth,angle=0]{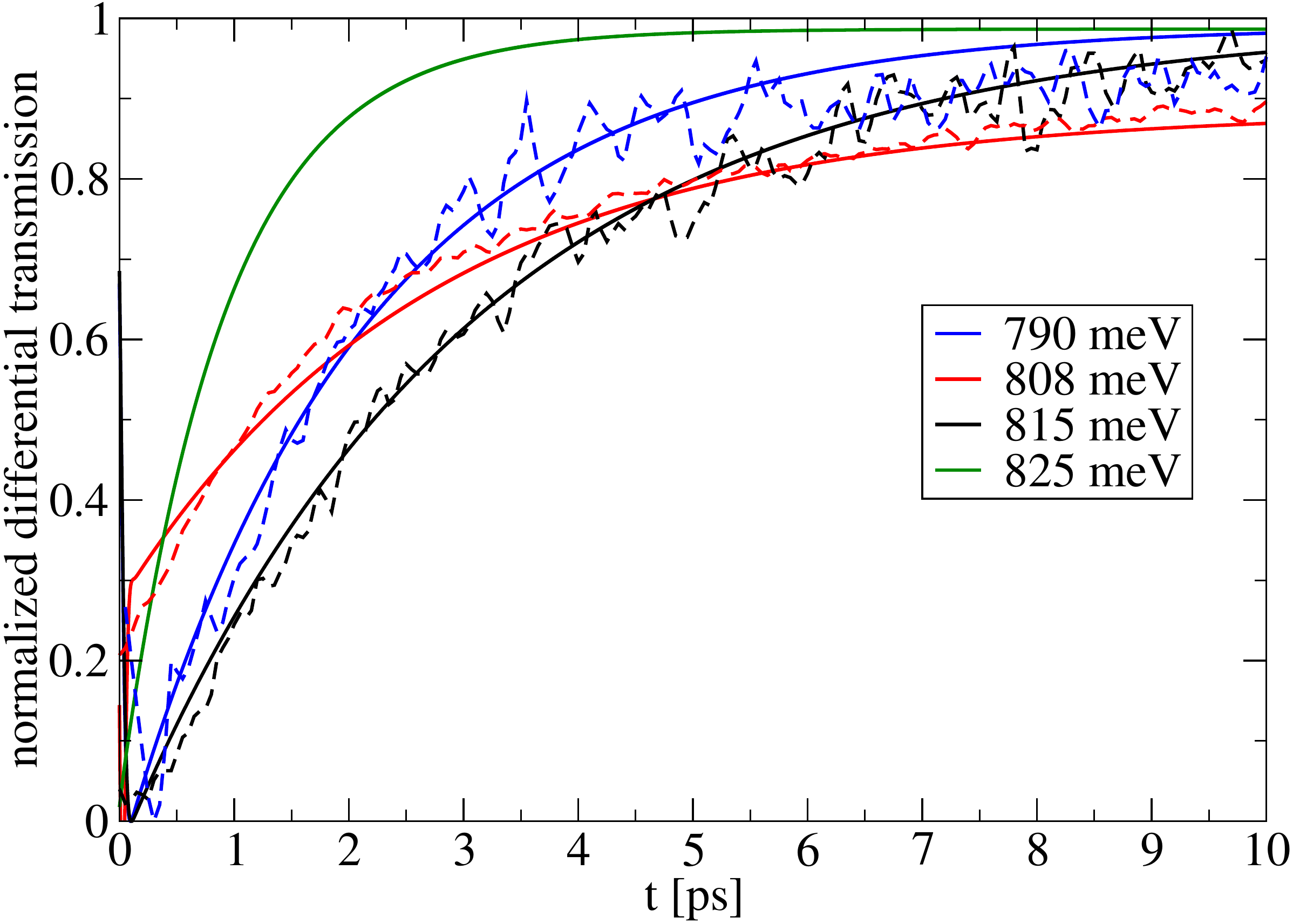}
    \caption{Comparison  of experimental (dashed line) and theoretical (solid line) differential pump probe traces for QDs with different s-shell energies within the inhomogeneous ensemble, identified via the probe energy of the respective QD ground state transition.
    \label{traces}}
  \end{center}
\end{figure}

In Fig.~\ref{traces} we compare the resulting differential pump probe time traces to the experimental results for QDs with different s-shell energies within the inhomogeneous ensemble.
For all QDs, the optical gain of the QD ground state is depleted, as shown by a sharp decrease of the differential pump probe signal during
the 100fs pulse, centered at t=0. Following the pulse, the gain recovers due to a refilling of the QD states via carrier-phonon and carrier-carrier scattering 
as discussed above.

Individual members of the QD ensemble couple differently to the iQW.
QDs with their first excited state near the quantum well bottom profit most from  tunnel coupling.
The situation with nearly optimal tunnel coupling via the hybridized p-shell (825 meV s-shell transition energy)  corresponds to Fig.~\ref{schematic2}a.
If the energy difference between ground state and quantum well exceeds the LO phonon energy by more than 10\%, 
scattering becomes increasingly inefficient (815 and 808 meV s-shell transition energy).
Therefore, within 20-30meV we find QDs that benefit substantially different from the tunnel coupling, as confirmed 
by the experimental results (dashed lines).
%
% In quantum dots with increasing confinement depth,  excited states become sucessively confined.
% %
% Here, scattering gets more efficient again, as the next excited state becomes 
% confined and reaches the phonon resonance with the quantum well bottom.

%For the QDs  with transition energies of 790 meV and 815 meV excellent agreement between theory and experiment is found. %For the 
%QDs in the center of the inhomogeneous distribution, at 808 meV, a somewhat larger deviation is observed, i.e., the %experimental time trace shows a faster gain recovery 
%than the theoretically predicted behavior. 
%This could be due to the fact that  the barrier states that refill the higher lying QW states
%are neglected in the calculations, 
%and this leads to a faster recovery. 

\begin{figure}[h!]
  \begin{center}
%   \hspace*{-2cm}%
    \includegraphics[width=0.5\textwidth,angle=0]{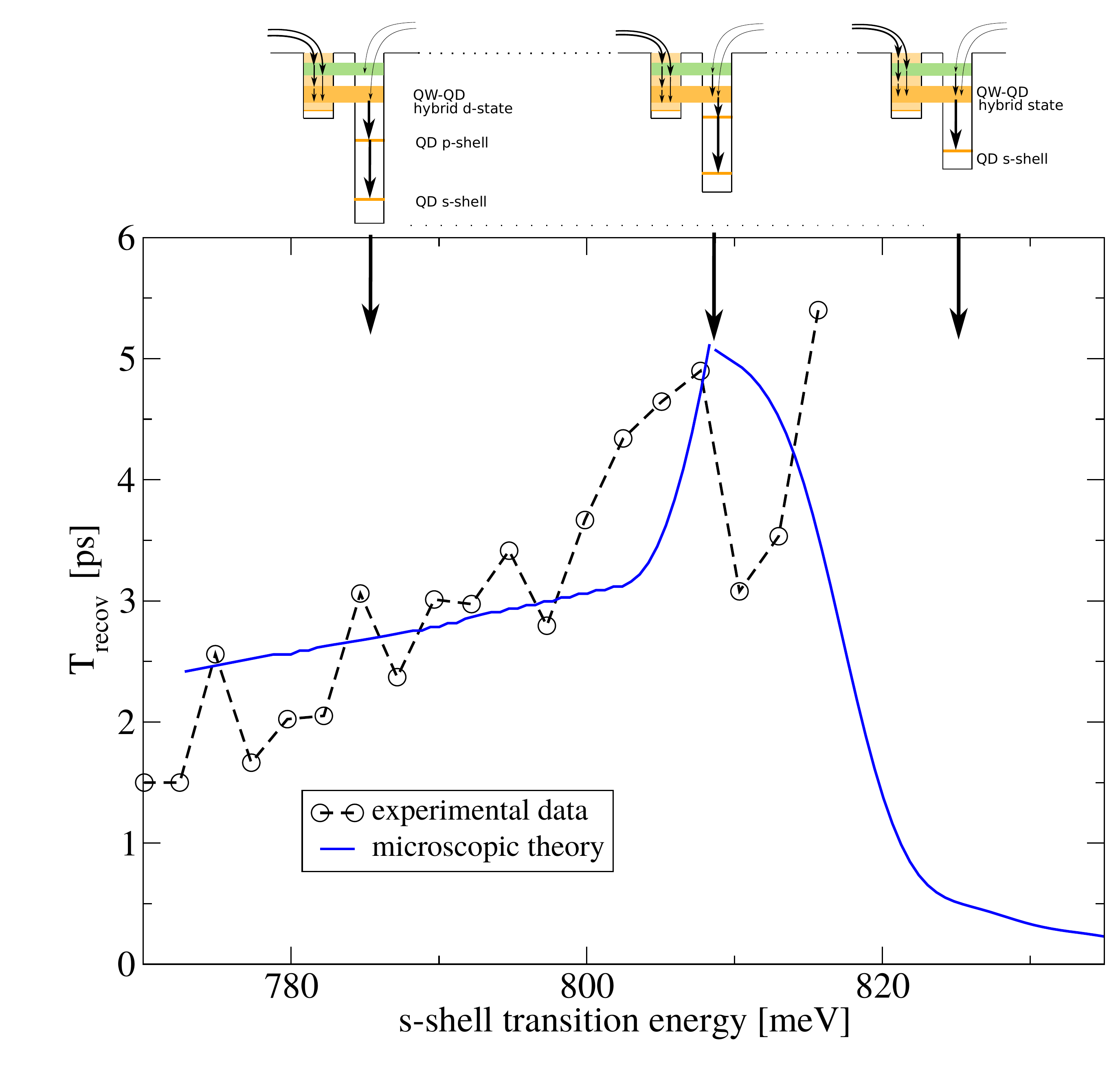}
    \caption{Gain recovery time for the QD electron s-shell population as a function of the s-shell transition energy in
comparison to experimental results. The sketches at the top illustrate the confinement situation in the particular energy range.
    \label{results}}
  \end{center}
\end{figure}

As expected, an approximately exponential recovery of the differential transmission is found in the time traces, that allows a fit via 
\begin{displaymath}
 \Delta T=\Delta T_0 (1-\exp{(-t/\tau_\text{recov})}) ~.
\end{displaymath}
Fig.~\ref{results} shows the resulting gain recovery time constants $\tau_\text{recov}$ for the QD electron s-shell population as a function of the s-shell transition energy.
% We can explain the experiment observations and additionally make predictions about QD states are that lie closer to the iQW continuum compared to those in current samples.
For energies larger than 820meV, we have a region where our theory predicts a recovery on the sub-ps timescale, due to 
the efficient coupling of the s-shell to the hybridized states via LO phonons.
If the energy difference between the bottom of the injector QW and the maximum of the QD distribution 
exceeds significantly the LO phonon energy of 36\,meV, which corresponds to s-shell energies below 820meV, QDs are hardly refilled by LO phonon scattering and 
gain recovery is mediated by Coulomb interactions. As this happens on a timescale of several picoseconds  \cite{Michael:18}, gain recovery times rise substantially.
If the QDs confinement deepens even more, the p-shell becomes gradually localized, while the hybridization takes place between QW and d-shell states.
Localization of the p-shell leads to:
\begin{enumerate}
 \item[a)] an increase in the efficiency of the Coulomb scattering from the p-shell to the s-shell  since the wave-function overlap between these two states increases,
 \item[b)]  scattering from the hybridized region into the p-shell being more efficient as the energetic separation between 
 QW and p-shell approaches the LO phonon energy. 
\end{enumerate}
The (counter intuitive) conclusion is that for the experimentally available transition energies, the gain recovery is  \emph{faster} for \emph{deeper} QDs, as seen in the low-energy region 
of Fig.~\ref{results}.

The behavior of interchanging regions with slower and faster recovery times will repeat itself for further lowering 
of the QD confinement energy. The recovery time increases, once the energetic distance between the highest localized QD state and the 
iQW exceeds the LO phonon energy. When the QD confinement is further increased to support an additional 
localized state, recovery will get faster again, as the carrier capture into the QD
will happen via this state, just as in the energetic window between 810meV and 780meV.
However, such QDs were not found in the present sample.

\section{Conclusion}
We have
% have described a detailed calculation, confirmed by experimental results of the dynamics governing carrier 
% injection efficiency from an injector QW to an inhomogeneous ensemble of QDs. The calculation and experiment are time 
% domain descriptions of the recovery of the lasing state following a perturbation by a short pulse. 
% We 
shown that the carrier scattering in TI-QD lasers is strongly influenced by the level alignment between QW and QD states. 
Our gain recovery results are directly related to the modulation capabilities of the laser 
and shed light on the significance of proper energy alignment in fast TI-QD lasers.
The most efficient carrier scattering is
found, if the first excited state hybridizes approximately at the bottom of the iQW conduction band.
For QDs with larger energy separation to the ground state, phonon scattering becomes increasingly inefficient
if the energetic separation exceeds the LO phonon energy.
With increasing confinement depth more QD states become localized, 
and provide efficient carrier scattering again, 
depending on the respective energetic separation of the highest localized state to the iQW bottom.
Hence, regions of efficient and inefficient efficiency alternate.
To optimze the modulation speed of TI-based laser devices, the inhomogeneously broadened sample 
as well as the iQW need to be carefully tuned to these conditions.

\section{Acknowledgment}
This work was partially supported by The Israel Science Foundation under grant number 460/21. I.K. acknowledges the support 
of the Rothschild Fellowship (The Yad Hanadiv Foundation), and of the Helen Diller Quantum Center at Technion.
M.L and F.J acknowledge supoort from the German Science Foundation (DFG) under grant number JA619/16-2 and CPU time at HLRN (Berlin/G\"ottingen).

% \bibliography{mybib2}
% \bibliographystyle{unsrtnat}

\end{document}